\documentclass[aps,prd,twocolumn,amsfonts,amssymb,amsmath,eqsecnum,nofootinbib,floatfix]{revtex4}
\usepackage[dvips]{graphicx}
\usepackage{psfrag}
\usepackage{dcolumn}
\begin{document}
\title{Estimate of Tilt Instability of Mesa-Beam and \\
       Gaussian-Beam Modes for Advanced LIGO \\ }
\author{Pavlin Savov}
\affiliation{Theoretical Astrophysics; California Institute of Technology, Pasadena, California}
\author{Sergey Vyatchanin}
\affiliation{MSURG LSC Group;  Physics Department, Moscow State University, Moscow, Russia}

\date{\today}

\def\lsim{\mathrel{\rlap{\lower4pt\hbox{\hskip1pt$\sim$}}
    \raise1pt\hbox{$<$}}}
\def\gsim{\mathrel{\rlap{\lower4pt\hbox{\hskip1pt$\sim$}}
    \raise1pt\hbox{$>$}}}

\begin{abstract}
Sidles and Sigg have shown that advanced LIGO interferometers will encounter a serious
{\it tilt instability}, in which symmetric tilts of the mirrors of an arm cavity cause the cavity's
light beam to slide sideways, so its radiation pressure exerts a torque that increases
the tilt.  Sidles and Sigg showed that the strength $T$ of this torque is 26.2 times greater for
advanced LIGO's baseline cavities --- nearly {\it flat} spherical mirrors
which support {\it Gaussian} beams (``FG'' cavities), than for nearly {\it concentric} spherical mirrors
which support Gaussian beams with the same diffraction losses as the baseline case --- ``CG''
cavities:  $T^{FG}/T^{CG} = 26.2$.
This has motivated a proposal to change the baseline design to nearly concentric, spherical mirrors.
In order to reduce thermoelastic noise in advanced LIGO, O'Shaughnessy and Thorne have
proposed replacing the spherical mirrors and their Gaussian beams by ``Mexican-Hat'' (MH) shaped mirrors
which support flat-topped, ``mesa'' shaped beams.  In this paper we compute the tilt-instability
 torque for advanced-LIGO cavities with nearly flat MH mirrors and mesa beams (``FM'' cavities) and
nearly concentric MH mirrors and mesa beams (``CM'' cavities), with the same diffraction losses
as in the baseline FG case.  We find that the relative sizes of the restoring torques are
$T^{CM}/T^{CG} = 0.91$, $T^{FM}/T^{CG} = 96$, $T^{FM}/T^{FG} = 3.67$.  Thus, the nearly
concentric MH mirrors have a weaker tilt instability than any other configuration.   Their thermoelastic
noise is the same as for nearly flat MH mirrors, and is much lower than for spherical mirrors.
\end{abstract}
\maketitle

\section{Introduction}

Thermoelastic noise is the dominant noise source in advanced-LIGO interferometers
with sapphire mirrors, at and somewhat below the frequency of optimal sensitivity \cite{tenoise1, tenoise2}.
O'Shaughnessy and Thorne \cite{mbi,mbis} have proposed lowering this thermoelastic
noise by flattening the cross-sectional profile of the arm cavities' light beams ---
i.e., by replacing the standard Gaussian-shaped beams by ``mesa''-shaped beams
(thick curves in Fig.~\ref{fig:modes} below). This can be achieved by
replacing LIGO's nearly flat, spherically shaped mirrors by mirrors that have a nearly
flat "Mexican-hat'' (MH) shape.  O'Shaughnessy, Strigin, and Vyatchanin \cite{mbi,mbis,osv} have shown
that the resulting reduction of thermoelastic noise power is about a factor 3, and with d'Ambrosio
and Thorne, they have carried out extensive studies which suggests that MH mirrors have no
serious disadvantages compared to spherical mirrors  \cite{mbi,mbis,osv,dA}.

\begin{figure}
\includegraphics[width=0.45\textwidth]{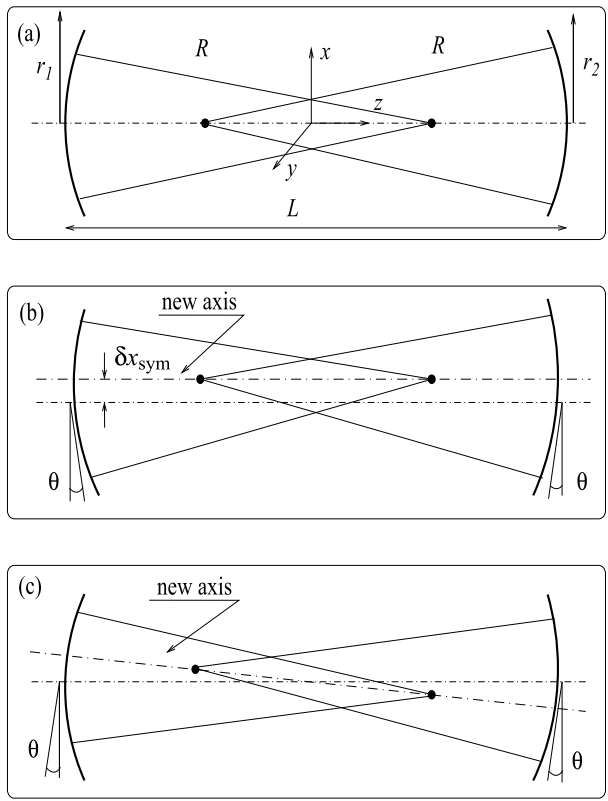}
\caption{\label{fig:tilt}FP resonator with (a) perfectly positioned spherical mirrors,
	(b) symmetrically tilted spherical mirrors and
	(c) antisymmetrically tilted spherical mirrors}
\end{figure}

Sidles and Sigg\cite{sigg,sidles} have recently rediscovered a {\it tilt instability} in Fabry-Perot (FP) cavities,
first pointed out by Braginsky and Manukin \cite{brag}, and they have shown that this instability is
a serious issue for advanced LIGO's arm cavities, because of their high circulating light power (about
800 kW) and resulting high light pressure.   In this instability, random forces cause the cavity's
mirrors to tilt in a symmetric way\footnote{Sidles and Sigg \cite{sigg,sidles} use the opposite convention from us for ``symmetric and ``antisymmetric'' tilt.} (Fig.~\ref{fig:tilt}b), and this tilt causes the light beam to slide sideways
in the cavity by the distance $\delta x_{\rm sym}$ shown in the figure, so its light pressure exerts a
torque $T$ on the mirrors that tries to increase their tilt.  [Sidles and Sigg also showed that, when
the mirrors are tilted in an antisymmetric way as in Fig.~\ref{fig:tilt}c, the resulting torque is
stabilizing rather than destabilizing.]  Sidles and Sigg analyzed the tilt instability, using geometric
arguments, for cavities with nearly {\it flat}, spherical mirrors and their {\it Gaussian} light beams (``FG''
cavities), and
also for nearly {\it concentric}, spherical mirrors and their Gaussian beams (``CG'' cavities).  [The mirrors
must be nearly flat or nearly concentric in order to make the light beams significantly larger than the
Fresnel diffraction size, $b=\sqrt{\lambda L/2\pi}$ with $\lambda$ the light's wavelength and $L$ the cavity
length; large beams are required to keep the thermoelastic noise small.]  Sidles and Sigg
found that the instability is much more severe for the baseline FG cavities than for CG cavities with the same
beam radii at the mirrors and thence the same diffraction losses. On this basis, the baseline design for advanced LIGO \cite{baseline} has been changed from FG cavities with nearly flat mirrors to CG cavities with nearly concentric mirrors.

Motivated by this Sidles-Sigg work, Thorne has proposed a mathematical way to design nearly concentric
MH mirrors that support mesa beams with precisely the same mesa-shaped
light-power distributions on the mirrors as for the original nearly flat
MH mirrors.  Thorne's mathematical construction is presented, along with some generalizations of it,
in a companion paper by Bondarescu \cite{mihai}.

In the present paper, we analyze the tilt instability for advanced-LIGO arm cavities with (i)
nearly flat MH
mirrors and their mesa beams (``FM'' cavities), and (ii) with Thorne's new nearly concentric MH
mirrors and their mesa beams (``CM'' cavities).
We employ first-order perturbation theory in our analysis, by contrast with the Sidles-Sigg geometric
techniques.   We compare the strength of the tilt's destabilizing torque $T$ for FG, FM, CG, and CM
cavities that have beam sizes chosen so they all have the same diffraction losses, about 20 ppm; and we explore
two choices for the radius of the mirror coating on the substrates: the baseline radius (14.7~cm), and
a larger coated radius (16~cm) used in the analysis of d'Ambrosio et.\ al.\ \cite{mbi,mbis} [their {\it fiducial} configuration].

In our numerical solutions to the eigenequation for the light's eigenmodes inside FM and CM
cavities, we discovered remarkable {\it duality relations} between cavities with axisymmetric
mirrors that deviate by an amount $H(r)$ from flatness, and cavities with mirrors that deviate by
$-H(r)$ from concentric spheres.  We verified these numerically discovered duality relations for
several different forms of $H(r)$, in addition to those of MH mirrors.  This motivated Chen and Savov, and independently Agresti and d'Ambrosio \cite{duality} to devise analytic proofs of our duality relations.  The duality relations provide a unique one-to-one mapping between the eigenstates and eigenvalues of the dual cavities --- a mapping that may be useful not only for advanced LIGO but in a variety of other applications of Fabry Perot cavities.

This paper is organized as follows.  In Sec.\ II, we use a first-order modal analysis of a Fabry-Perot
 cavity to derive a general formula for the torque exerted on the mirrors when the cavity is perturbed, in terms of as-yet unknown mode coupling coefficients $\alpha_k$ and mode-overlap integrals $I_k$. In Sec.\ III, we use first-order perturbation theory of Gaussian-beam (FG and CG) cavities to derive analytical formulas for $\alpha_k$ and $I_k$, and thence
for the tilt-induced torque $T$ in the FG and CG cases, and we show that our formula for the torque is  equivalent to that of Sidles and Sigg \cite{sigg,sidles}.  In Sec.\ IV we use first-order perturbation theory to derive formulas for the coupling coefficients $\alpha_k$, and thence for the torque $T$, in terms of  a cavity's eigenvalues and mode-overlap integrals $I_k$.  In Sec.\ V we present our numerical
results for the modes and their eigenvalues for FM, CM, FG, and CG cavities, and we discuss the
duality relations between the nearly flat and nearly concentric cases.  Finally, in Sec.\ VI, we
combine the numerical results of Sec.\ V with the formulas of Secs. III and IV, to deduce the
tilt-induced torque for our four cavity designs --- using two sets of parameters: those for cavities
with advanced-LIGO baseline mirror radii, and those for d'Ambrosio et.\ al.'s slightly larger mirrors (``fiducial'' configurations). We present a brief conclusion in Sec.\ VII. For the readers interested in our numerical implementation of the eigenvalue problem, we include an Appendix where we sketch details of our computational work.

The results presented in this paper are based on previous work on nearly flat
configurations by S. Vyatchanin \cite{vy} (some errors in this paper are
corrected here) combined with recent analyses of
nearly concentric cavities by P. Savov. An analytical proof of the duality
relation between nearly flat and nearly confocal resonators by P. Savov and Y.
Chen, and independently by E. D'Ambrosio and J. Agresti, will be provided in a
companion paper \cite{duality}.

\section{Main formulas}

The light inside LIGO arm cavities is well-described by the laws of diffraction optics in the paraxial approximation.
The eigenvalue problem in this approximation for a half trip through a cavity with two identical axisymmetric mirrors can be written as
\begin{equation}
\int G(\vec {r}_1, \vec {r}_2)\, u(\vec {r}_2)\, d^2 \vec {r}_2=\lambda\, u(\vec {r}_1).
\end{equation}
In the above equation $u(\vec r)$ is an eigenmode of the cavity and $\lambda$ is the corresponding eigenvalue.
The eigenmode represents the state of the light (the electric field) on the surface of a mirror.

For advanced LIGO diffraction losses will be very small (about 10 ppm for each half trip), so it is an excellent idealization to ignore the losses and idealize the mirrors as infinite in radius. Then, $|\lambda|=1$, $G$ is a unitary operator, and its eigenvectors form a complete set. Each eigenmode $u_{nm}$ and the corresponding eigenvalue $\lambda_{nm}$ are labeled by two ({\it quantum}) numbers --- radial (or principle) number $n = 0, 1, \ldots$ and angular (or azimuthal) number $m = 0, 1, \ldots$. All modes with angular number $m=0$ are axisymmetric (no angular dependence), $m=1$ are dipolar, $m=2$ are quadrupole, etc.:
\begin{equation}
\label{angulardep}
u_{nm}\propto e^{-im\varphi}.
\end{equation}
The eigenmodes are normalized and orthogonal to each other according to the following definition:
\begin{equation}
\label{orthonormal}
\int u_{n_1m_1}(\vec r)\,u_{n_2m_2}^*(\vec r)\,d^2 \vec r = \delta_{n_1n_2}\, \delta_{m_1m_2}\, .
\end{equation}
We will use this set of eigenvectors as a basis for expanding the eigenmodes
of cavities with tilted mirrors.
The radial coordinate $r$ is dimensionless and  measured in units of the Fresnel diffraction size
\begin{equation} 
b=\sqrt{L\lambda/2\pi}.
\end{equation}

When the mirrors of a FP cavity are tilted in a symmetric way (as in Fig.~\ref{fig:tilt}b), the cavity's fundamental mode $u_{00}(\vec r)$ is transformed into the fundamental mode $\tilde u_{00}(\vec r)$ of the perturbed cavity. The torque acting on the mirrors when the light is in this mode and has power $P$ is

\begin{equation}
\label{torque0}
T= \frac{2Pb}{c} \int \big|\tilde u_{00}(\vec r)\big|^2 r\cos\varphi\,
        d^2 \, \vec r\,.
\end{equation}

The new fundamental mode can be expanded over the set of orthonormal modes $\{u_{nm}(\vec r)\}$ of the unperturbed cavity 
\begin{equation}
\label{series0}
\tilde u_{00}(\vec r) = u_{00}(\vec r) + \sum_{n,m}\alpha_{nm}u_{nm}(\vec r).
\end{equation}
In this paper, we study effects only to first order in the perturbation. That is why the coefficient in front of $\tilde u_{00}(\vec r)$, in Eq.~(\ref{series0}), is unity.

By substituting Eq.~(\ref{series0}) into Eq.~(\ref{torque0}) and using the angular dependence of the eigenmodes Eq.~(\ref{angulardep}), we conclude that only the dipolar eigenmodes ($m=1$) contribute to the net torque and more specifically their part proportional to $\cos(\varphi)$. Thus, for our purposes of calculating the torque, we will assume $u_{n1}\propto \cos(\varphi)$. Since the only modes we use from now on are the fundamental mode $u_{00}$ and all dipolar modes $u_{n1}$, in order to  simplify notation, we collapse the indices into one labeling index
\begin{equation}
\label{indexing}
k=n+m.
\end{equation}
Thus the fundamental mode becomes $u_0$, the first dipolar mode becomes $u_1$ (corresponding to the old notation $u_{01}$) and so on. When necessary, we will use  the conventional notation with two labeling indices.

We will study the effects of tilt only to first order in the tilt angle $\theta$, so for our purposes we use the following expansion of the perturbed eigenmode:

\begin{eqnarray}
\label{series}
\tilde u_{0}(\vec r)& = &
	u_{0}(\vec r) + \sum_{k=1} \alpha_{k}u_{k}(\vec r) ,\\
\label{u0}
u_{0}(\vec r) &=& \frac{u_{0}(r)}{\sqrt{2\pi}},\\
\label{uk}
u_{k}(\vec r) &=& \frac{u_{k}(r)\cos \varphi}{\sqrt\pi},\\
\label{norm0}
\int_0^\infty \big[u_{0}(r)\big]^2\, r\, dr&=&1, \\
\label{normk}
\int_0^\infty \big[u_{k}(r)\big]^2\, r\, dr&=&1, \qquad k=1,2, \ldots.
\end{eqnarray}
In the above equations, $u_{k}(\vec r)$  are the dipolar modes on the surface
of a mirror; $u_{k}(r)$ are their parts depending only on the radial
coordinate $r$; all $u_{k}(r)$ are dimensionless and normalized as shown above
[cf. Eq.~(\ref{orthonormal}) with $m=0,1, k=n+m$]; and $\alpha_k$ are
dimensionless coupling constants, proportional to the mirrors' tilt angle
$\theta$, which we will evaluate in Sec.~III for Gaussian (FG and CG) beams
and in Sec.~IV for mesa (FM and CM) beams.
In general,
$u_{k}(\vec r)$ are complex fields, but since the mirror surfaces coincide
with the beam's wave front, up to an overall complex phase which we chose to
be zero, they are real fields.

Now we can calculate the torque that the cavity's light exerts on each mirror:

\begin{eqnarray}
\label{T}
T& = & \frac{2Pb}{c} \int [\tilde u_{0}(\vec r)\big]^2
	r\cos\varphi\,  r\, dr\, d\varphi
	= \nonumber\\
 & = & \frac{2Pb}{c} \,2 \sum_{k=1} \alpha_{k}
	\int u_{0}(\vec r)u_{k}(\vec r)\,
	r\cos\varphi\,  r\, dr\, d\varphi \,, \nonumber
\end{eqnarray}
where we have used Eq.~(\ref{series}). By inserting Eqs.~(\ref{u0}) and (\ref{uk}), we obtain the following formulas for the tilt-induced torque to first order in $\alpha_k$ (first order in $\theta$):
\begin{eqnarray}
\label{torque}
T & = & \frac{2\sqrt 2\, Pb, }{c}\times \sum_{k=1}\alpha_{k}\, I_{k},\\
\label{integral}
I_{k} &=& \int u_{0}(r)u_{k}(r)\,r^2\, dr\, .
\end{eqnarray}
These formulas are valid for any FP cavity and in particular for FG, FM, CG, and CM cavities that interest us (of course, the modes $u_{k}(r)$ are different for different cavities).

In the sections below, we calculate the values for the coupling constants $\alpha_{k}$ and the overlap integrals $I_k$ for our four types of cavities. Our analysis for conventional spherical mirrors (FG and CG; Sec. III) is entirely analytical, whereas for any generic mirror shape, and MH mirrors in particular (FM and CM; Sec IV), numerical treatment is required. We will test our numerical methods by applying them to FG and CG cavities and comparing with the analytical results.

\section{Gaussian-Beam (FG and CG) Cavities}

We consider a cavity with identical spherical mirrors. We are interested in a symmetric tilt of the two mirrors by a small angle $\theta$ as shown on Fig.~\ref{fig:tilt}b. In this case, the axis of the new mode $\tilde u_{0}(\vec r)$ is displaced by a small distance $\delta x_{\rm sym}$, but is still parallel to the old axis. The field distribution on each mirror will be unchanged, but shifted by $\delta x_{\rm sym}$.

Spherical cavities have been studied thoroughly (see e.g \cite{siegman1}); their fundamental modes are the well-known Gauss-Laguerre modes (called in this paper FG and CG modes). We will use these modes derive analytical formulas for $\alpha_{k}$ and $I_k$. The main axisymmetric and dipolar modes [$u_{0}(r)$ and $u_{1}(r)$] are given by (see e.g. \cite{siegman2}):

\begin{eqnarray}
\label{gauss0}
u_{0}^{G}(r) &=& \frac{\sqrt 2}{r_0} e^{-r^2/2r_0^2},\quad
	\\
\label{gauss1}
u_{1}^{G}(r) &=& \frac{\sqrt 2 \, r}{r_0^2} e^{-r^2/2r_0^2},\quad
        \\
\label{gaussr0}
r_0 &=& \frac{1}{(1-g^2)^{1/4}}.
\end{eqnarray}

Here  $r$ is the dimensionless radial coordinate (measured in units of $b$),  $r_0$ is the dimensionless radius of the beam at the mirrors' surface (also in units of $b$), $g=1-L/R$ is the so-called $g$-parameter of the cavity, $L$ is the distance between the mirrors, and $R$ is the mirrors' radius of curvature (Fig.~\ref{fig:tilt}a). (The intensity on the mirror is proportional to $e^{-r^2/r_0^2}$.)

For spherical mirrors the displacement of the optic axis $\delta x_{\rm sym}$ is (see Fig.~\ref{fig:tilt}b):
\begin{equation}
\delta x_{\rm sym} \simeq \frac{R\theta}{b} = \frac{L\, \theta}{b(1-g)}.
\end{equation}

Next, we write down the main mode $\tilde u_{0}^{G}$ of the FP resonator with tilted mirrors and expand it to first order in $\delta x_{\rm sym}$:
\begin{eqnarray}
\tilde u_{0}^{G} (\vec r) &=& \frac{e^{-r_{\delta x}^2/2r_0^2}}{\sqrt{\pi}},\\
r^2_{\delta x}&=&\Big(r\cos \varphi- \delta x_{\rm sym}\Big)^2+
            r^2\sin^2\varphi,     \\
\tilde u_{0}^{G} (\vec r) & = & u_{0}^{G}(\vec r)\left(1+ \frac{r\, \delta x_{\rm sym}\,
       \cos \varphi}{r_0^2}\right)\nonumber \\
&=& u_{0}^{G}(\vec r) + \frac{ \delta x_{\rm sym}\,
       \cos \varphi}{\sqrt{2\pi}r_0} \, u_{1}^{G}(r)\nonumber \\
&=& u_{0}^{G}(\vec r) +
     \underbrace{\frac{\delta x_{\rm sym}}{\sqrt{2}r_0}}_{\alpha_1^{G}}
     \underbrace{
     \frac{u_{1}^{G}(r)\, \cos\varphi}{\sqrt \pi}}_{u_{1}^{G}(\vec r).
     }
\end{eqnarray}
As we can see, the only nonzero coupling constant is $\alpha_{1}^{G}$
\begin{equation}
\label{alpha1g}
\alpha_{1}^{G} = \frac{L\,\theta\, (1+g)^{1/4}}{\sqrt{2}\,b\,(1-g)^{3/4}}.
\end{equation}

From Eqs.~(\ref{integral}), (\ref{gauss0}), (\ref{gauss1}), and (\ref{gaussr0}), we can easily calculate the only overlap integral we need for Gaussian beams:

\begin{eqnarray}
\label{ig}
I_1^{G} &=& \int_0^\infty u_{0}^{G}(r)u_{1}^{G}(r)\,r^2\, dr=\nonumber\\
&=&r_0 = \frac{1}{(1-g^2)^{1/4}}.
\end{eqnarray}
Substituting into Eq.~(\ref{T}) along with Eq.~(\ref{alpha1g}) we derive a final expression for the torque:
\begin{equation}
\label{torqueg}
T^G=\frac{2PL}{c}\frac{\theta}{(1-g)}.
\end{equation}

This result, derived by a modal analysis, is in complete agreement with the result of the Sidles-Sigg geometrical analysis in its {\it long-cavity} limit (Section~5 of \cite{sidles}). In their notation, the torque for the unstable configuration is
\begin{equation}
T^G=-k_{-}\,\theta=\frac{2PL}{c}\frac{\theta}{(1-g)},
\end{equation}
where $=k_{-}$ is the negative eigenvalue of {\it a torsional stiffness matrix} (Eq.~(23) of Section~5 in \cite{sidles}). (Note that negative eigenvalues in the Sidles-Sigg analysis are associated with unstable configurations --- the subject of interest in this paper.) Our perturbation method gives the exact result (to first order in $\theta$) for spherical mirrors, because the only contribution to the torque is from the lowest dipolar mode $u_1$. This is a property only for spherical mirrors and their Gaussian beams. As we'll see in the following sections, for any generic mirror shapes, we have to calculate the contribution from all higher dipolar modes.

\section {Mesa-Beam (FM and CM) Cavities: Analytical Formulas}

\paragraph{Perfectly positioned mirrors (Fig.~\ref{fig:tilt}a).}
For any cavity with axisymmetric mirrors, and in particular MH mirrors, the  main axisymmetric mode $u_{0}(\vec r)$  and all dipolar  modes $u_{k}(\vec r)$ satisfy the integral
eigenequations
\begin{eqnarray}
\label{eigenmesa0}
\int G(\vec r_1, \vec r_2)\, u_{0}(\vec r_2)\, d^2 \vec r_2 &=&
	\lambda_0\, u_{0}(\vec r_1),\\
\label{eigenmesa1}
\int G(\vec r_1, \vec r_2)\,u_{k}(\vec r_1)\, d^2 \vec r_1 &=&
	\lambda_{k}\, u_{k}(\vec r_2),
\end{eqnarray}
where $G, u_{0}, u_{k}, \vec{r_1}, \vec{r_2}$ are all dimensionless and the eigenvalue of the $k^{\rm th}$ dipolar mode $u_{k}$ is $\lambda_{k}$.

In the paraxial approximation, the kernel of the operator $G$ is the following (up to a trivial factor of $e^{ikL}$ due to phase accumulation along the arm length $L$, which we omit, thereby fixing a common overall phase factor in all the $\lambda_k$):
\begin{eqnarray}
\label{G0a}
G(\vec r_1, \vec r_2) &=& \frac{-i}{2\pi}
	\exp \left[i\left(\frac{(\vec r_1-\vec r_2)^2}{2} -
		 h_1(\vec r_1) -h_1(\vec r_2) \right) \right], \nonumber\\
h_{1,2}(\vec r) & = & k H_{1,2}(\vec r), \qquad  k=\frac{2\pi}{\lambda}.
\end{eqnarray}
Here $H_1(\vec r_1)$ and $H_2(\vec r_2)$ are the physical deviations of the mirrors' surfaces from a plane surface, which we assume to be the same, $H_1(\vec r_1) = H_2(\vec r_2)$ (identical mirrors).

\paragraph{Symmetrically tilted mirrors (Fig.~\ref{fig:tilt}b).}
The tilt is equivalent to small deviations of each mirror's position from the unperturbed one:
\begin{subequations}
\begin{eqnarray}
\delta h_1 = k b\, r_1\,\cos \varphi_1\  \theta && \mbox{(left mirror)}\\
\delta h_2 = k b\, r_2\, \cos\varphi_2 \ \theta && \mbox{(right mirror)}.
\end{eqnarray}
\end{subequations}
These tilts induce a coupling of all the dipolar modes $u_1, u_2,\ldots$ into the cavity's fundamental mode $\tilde u_0$, as shown in Eq.~(\ref{series}), though (as our numerical work will show) the coupling for the first dipolar mode is far greater than the others $\alpha_1 I_1 \gg \alpha_k I_k$ for $k=2,3,\ldots$.

For simplicity, we will show the analysis only for the first dipolar mode $u_1$ ($u_{01}$ in the conventional notation). The generalization for the higher dipolar modes is trivially obtained by by replacing the subscript $1$ by the desired dipolar mode's subscript $k$

The eigenvalue of the fundamental mode of the perturbed cavity $\tilde \lambda_0$ will slightly differ from $\lambda_0$:
$\tilde \lambda_0 = \lambda_0 +\Delta$. Thus, we have the following integral
eigenequation for $\tilde u_0(\vec r)$
\begin{eqnarray}
\label{tilteigena}
&&(\lambda_0+\Delta)[ u_0(\vec r_1) + \alpha_1 u_1(\vec r_1)]
        =\nonumber\\
&&= \int G (\vec r_1, \vec r_2)
	[ 1 - i\,\delta h_1(\vec r_1)-i\delta h_2( \vec r_2)]
                        \nonumber\\
&&\qquad       \times
        [u_0(\vec r_2) + \alpha_1 u_1(\vec r_2)]\,d^2\vec r_2.
\end{eqnarray}
This equation can be simplified by use of the eigenequation of the original unperturbed system (\ref{eigenmesa0}):
\begin{eqnarray}
&&\Delta u_0 (\vec r_1) +
        (\lambda_0 + \Delta-\lambda_1)\alpha_1 u_1 (\vec r_1)
        =\nonumber\\
\label{tilteigen}
&&= -   i\int G (\vec r_1, \vec r_2)
	[ \,\delta h_1(\vec r_1)+\delta h_2( \vec r_2)] \nonumber
                        \\
&&\qquad  \quad     \times
        [u_0 (\vec r) + \alpha_1 u_1 (\vec r)]\,d^2\vec r_2.
\end{eqnarray}
Multiplying Eq.~(\ref{tilteigen}) by $u_0 (\vec r_1)$ and integrating
over $d^2\vec r_1$, one can find that the correction $\Delta$ to the eigenvalue $\lambda_0 $ has second order of smallness, $\Delta \sim \theta^2$, so below we set $\Delta=0$.

Multiplying Eq.~(\ref{tilteigen}) by $u_1 (\vec r_1)$ and integrating over $d^2\vec r_1$, one can find $\alpha_1$:
\begin{eqnarray}
\big(\lambda_0 -\lambda_1 \big)\alpha_1 &=& -i\big(\lambda_0+\lambda_1\big) \nonumber
        \\
&\times & \int u_0(\vec r_1)\, u_1(\vec r_1)\, \delta h_1(\vec r_1)\,
                d^2\vec r_1,\nonumber
\end{eqnarray}
so
\begin{eqnarray}
\alpha_1 &=&  -\frac{ikb\, \theta\,
        \big(\lambda_0+\lambda_1\big)}{\sqrt 2 \big(\lambda_0-\lambda_1\big)}\underbrace{
        \int_0^\infty u_0(r) \, u_1(r)\, r^2\, dr}_{I_1}
        \nonumber\\
\label{alpha1m}
&=& -\frac{iL\,I_1 \theta (\lambda_0+\lambda_1)}{\sqrt 2 b(\lambda_0-\lambda_1)}.
\end{eqnarray}

Similarly for the higher dipolar modes
\begin{equation}
\label{alphak}
\alpha_k= -\frac{iL\,I_k \theta (\lambda_0+\lambda_k)}{\sqrt 2 b(\lambda_0-\lambda_k)}.
\end{equation}

In order to calculate the numerical value of $\alpha_k$, we must solve the eigenequations (\ref{eigenmesa0}) and (\ref{eigenmesa1}) numerically for the eigenvalues $\lambda_0$, $\lambda_k$ and the corresponding eigenfunctions  $u_0(r)$, $u_k(r)$ (see Appendix~\ref{A} for details). The value of the integral $I_k$ can be calculated numerically from Eq.~(\ref{integral}).

Note that the formulas in this section are valid for any resonators with symmetric mirrors $H_1(r_1)=H_2(r_2)$ and very low diffraction losses, not just for mesa-beam resonators.

\section{Numerical Solutions of Eigenequations}

\begin{figure}
\includegraphics[width=0.45\textwidth]{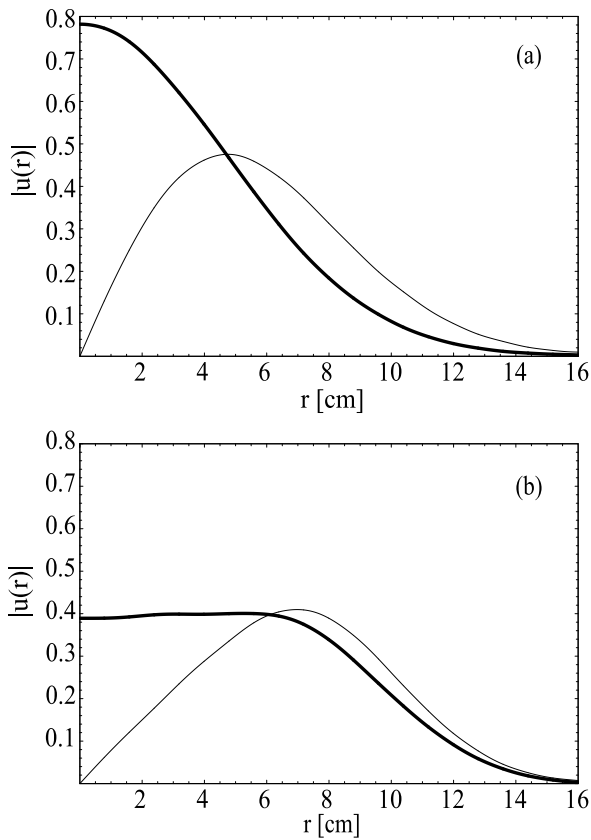}
\caption{\label{fig:modes} Fundamental modes $u_0(r)$ (thick curves) and first dipolar modes $u_1(r)$ (thin curves) at mirrors' surfaces for (a) FG and CG cavities, and (b) FM and CM cavities. The modes are dimensionless and normalized according to Eqs.~(\ref{norm0}), and(\ref{normk}). We have used the fiducial cavity parameters of d'Ambrosio et.\ al.: Eqs.~(2) of Sec.~IVA of \cite{mbi} and Sec.~IIIA of \cite{mbis}}.
\end{figure}

\begin{figure}
\includegraphics[width=0.45\textwidth]{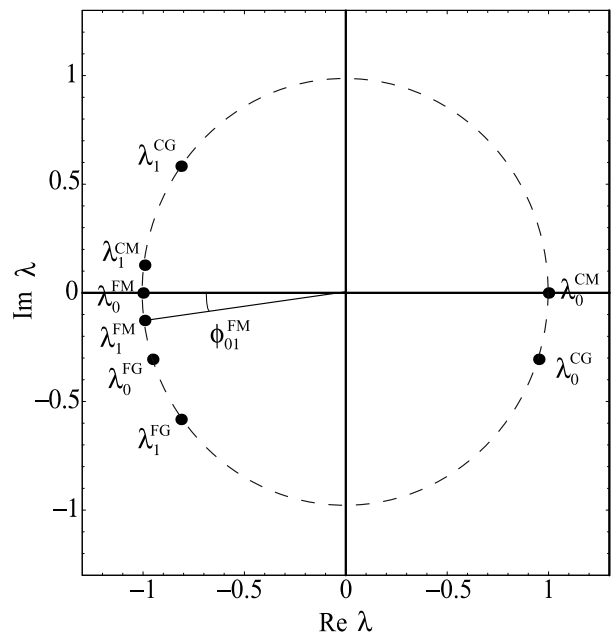}
\caption{\label{fig:vals} Eigenvalue spectrum in the complex plain. Note that all eigenvalues satisfy the duality relation, Eq.~(\ref{valsduality}) ($n=0,\; m=0$ for $\lambda_0$, and $n=0,\; m=1$ for $\lambda_1$); see also \cite{duality}.}
\end{figure}

We have solved the eigenequations (\ref{eigenmesa0}) and (\ref{eigenmesa1}) numerically using the scheme described in Appendix~\ref{A}, for our four cavity configurations: FG, CG, FM, and CM. Recall that our nearly flat and nearly concentric cavities were chosen such that the intensity $u_0(r)^2$ and therefore $u_0(r)$ at the mirrors' surfaces are identical (FG and CG are the same and FM and CM are the same). We have found numerically for FM and CM (mesa beams) and for FG and CG (Gaussian beams, Sec.~III) that $u_k(r)$ is also the same for the nearly concentric and nearly flat cases. The eigenfunctions $u_0$ and $u_1$ are shown in Fig.~\ref{fig:modes}. The eigenvalues, by contrast, are different for nearly flat and nearly concentric cavities, so we have four sets of eigenvalues (FG, CG, FM, CM), depicted in Fig.~\ref{fig:vals}.

In our numerical solutions to the eigenequations (\ref{eigenmesa0}) and (\ref{eigenmesa1}), we found an interesting duality relation between nearly flat and nearly concentric configurations. This duality relation is satisfied for any generic mirror shape that satisfies the paraxial approximation. To within numerical error of less than 0.05 per cent, we found that a nearly concentric cavity, which has the same intensity profile as a nearly flat configuration, also has the same mirror-shape correction as the nearly flat cavity, but with opposite sign:
\begin{equation}
\delta h^{C}(r)=-\delta h^{F}(r)\,.
\end{equation}
Here $\delta h^{C}(r)$ is the deviation from concentric spherical shape, and $\delta h^{F}(r)$ is the deviation from flat shape.
We also found, numerically, a unique mapping between the eigenvalues of these dual configurations:
\begin{equation}
\label{valsduality}
\lambda_{nm}^{C}=(-1)^{m+1}(\lambda_{nm}^{F})^*\,,
\end{equation}
for any pair of integers $n,m=0,1,2,\ldots$.
In addition, all higher modes have the same intensity profiles at the mirrors' surfaces as their counterparts
\begin{equation}
|u_{nm}^C|^2=|u_{nm}^F|^2
\end{equation}
for any integer $n,m=0, 1, \ldots$.

Remarkably, our numerical calculations showed that these relations hold not just for mesa-beam cavities, but for all stable cavities that we explored (all mirror shapes  $\delta h^{\rm C,F}$, including cavities in which the deviations $\delta h^{\rm C,F}$ from  concentric spherical and flat shapes are large --- so large as the paraxial approximation is valid).

This has led us to conjecture a {\it duality} relation between symmetric cavities with axisymmetric mirrors: for any two such cavities, A and B, with
\begin{equation}
\label{AB}
h^{\rm A}(r) + h^{\rm B}(r) = \frac{r^2}{L}
\end{equation}
there exists a one-to-one correspondence between their eigenstates: they all have the same intensity profiles at the mirrors, while
\begin{equation}
\label{conjecture}
\lambda_{nm}^{\rm A}=(-1)^{m+1}(\lambda_{nm}^{\rm B})^*\,.
\end{equation}

Chen and Savov, and independently Agresti and d'Ambrosio \cite{duality} have verified this conjecture analytically.

\section{Strength of the Tilt Instability for FG, CG, FM, and CM cavities}

We now have all the tools we need to compute the tilt-induced torque $T$ on the cavity's mirrors, for FG, CG, FM, and CM configurations. We shall evaluate $T$ for two sets of cavity parameters: the fiducial parameters used by d'Ambrosio et.\ al. \cite{mbi, mbis, osv} and the advanced-LIGO baseline parameters (Table~1 in \cite{baseline}).

The set of parameters for the fiducial cavity (see Sec.~IV\,A(2) of \cite{mbi} and Sec.~III\,A of \cite{mbis}) is:

$L=4$~km --- the length of the cavity.

$\lambda=1064$~nm --- the wavelength of the laser beam.

$k=2\pi/\lambda$ --- the wave number associated with $\lambda$.

$b=\sqrt{L\lambda/2\pi}= 2.603$~cm --- the natural diffraction length scale (Fresnel length).

$r_{max}=16$~cm --- the radius of the mirrors' coated surfaces.

$g^{FG}=0.952$ --- the $g$-factor for the fiducial FG resonator (corresponding mirror radius of curvature $R=83.33$~km).

$g^{CG}=-0.952$ --- the $g$-factor for the fiducial CG resonator (corresponding mirror radius of curvature $R=2.05$~km).

$r_0=b/(1-g^2)^{1/4}=4.7$~cm --- the radius of the FG and CG beams at the mirrors.

$D=4b=10.4$~cm --- the radius parameter of the FM and CM beams at the mirrors (see Sec.~II\,A and Sec.~IV\,A(2) of \cite{mbi}).

The above beam radii were chosen so as to make the diffraction losses be about 20 parts per million (ppm). More specifically, they are 23 ppm for the FG and CG beams and 19 ppm for the FM and CM beams.\footnote{We have deduced these diffraction
losses from our numerical solutions of the cavity's eigenequation.}

From Eqs.~(\ref{alpha1g}), (\ref{ig}), and (\ref{torque}) we can calculate the integral $I^G_1$, the coupling constant $\alpha^G_1$, and the torque $T^G$ for the FG and CG cavities. Our results are shown in the second and third column of Table~\ref{tab:res1}.

\begin{table}
\caption{\label{tab:res1}Comparison Between Analytical and Numerical Results for FG and CG Cavities; $\alpha_1$ is measured in units of $(\theta/10^{-8})$ and $T$ is in units of $(Pb/c)(\theta/10^{-8})$}
\begin{ruledtabular}
\begin{tabular}{lllll}
 &\multicolumn{2}{c}{Analytical}&\multicolumn{2}{c}{Numerical}\\
             &\multicolumn{1}{c}{FG}&\multicolumn{1}{c}{CG}&\multicolumn{1}{c}{FG}&\multicolumn{1}{c}{CG}   \\
\hline
   $I_1$     &1.8075  &1.8075    &1.8073  &1.8073  \\
   $\alpha_1$&0.012526&0.00030802&0.012525&0.00030799  \\
   $T$       &0.064038&0.0015747 &0.064023&0.0015743  \\
\end{tabular}
\end{ruledtabular}
\end{table}


We  have already established an agreement between our analytically derived results using the modal analysis described in Section III and the Sidles-Sigg results derived from geometric considerations \cite{sidles}. We can also test the numerical first-order perturbation methods that we developed for arbitrary mirror shapes by applying them to our FG and CG cavities. By substituting our numerical results for $u_0^{FG}=u_0^{CG}$, $u_1^{FG}=u_1^{CG}$, $\lambda_0^{FG}$, $\lambda_0^{CG}$, $\lambda_1^{FG}$, and $\lambda_1^{CG}$ into Eqs.~(\ref{ig}), (\ref{alpha1m}), and (\ref{torque}), we calculate the results shown in the last two columns of Table~\ref{tab:res1}. These numerical results all agree with our analytical results to within 0.05 per cent, thus validating our numerical methods.

As was found by Sidles and Sigg, the CG configuration is significantly less unstable than its nearly flat counterpart FG. The analytical analysis (first two columns in Table~\ref{tab:res1} predicts
\begin{equation}
\frac{T^{FG}}{T^{CG}}=\frac{1+g^{FG}}{1-g^{FG}}=\frac{R^{FG}}{R^{CG}}= 40.667,
\end{equation}
which is in agreement with the numerical result 40.667  (last two columns).

From the modal analysis applied to FG and CG cavities [Eqs.~(\ref{alpha1m}) and (\ref{torque})], we deduce that, aside from factors that are the same for FG and CG,
\begin{equation}
\label{torquemodal}
T^{G} \propto i \,\frac{\lambda^{G}_0+\lambda^{G}_1}{\lambda^{G}_0-\lambda^{G}_1}={\rm cotan}\left(\frac{\phi^{G}_{01}}{2}\right).
\end{equation}
Here, the equality holds because   $|\lambda| = 1$ for all modes (negligible diffraction losses) and $\phi^{G}_{01}$ is the phase separation between $\lambda^{G}_0$ and $\lambda^{G}_1$, i.e. the argument of  $\lambda^{G}_0/\lambda^{G}_1$ (in Fig.~\ref{fig:vals} we show $\phi^{FM}_{01}$ for the FM cavity). Thus, Eq.~(\ref{torquemodal}) is governed by the phase separation of the eigenvalues $\lambda^{G}_0$ and $\lambda^{G}_1$ . As Fig.~\ref{fig:vals} shows the two eigenvalues for the FG configuration are very close to each other so ${\rm cotan}(\phi^{FG}_{01}/2) \gg 1$, whereas the phase separation of the eigenvalues for the CG configuration is close to $\pi$ so ${\rm cotan}(\phi^{CG}_{01}/2) \ll 1$. This explains why $T^{FG} \gg T^{CG}$.

Similarly to the above Gaussian analysis, we use our numerical results to compute the torques $T^{FM}$ and $T^{CM}$ for FM and CM cavities respectively. In this case, we must include the contributions from higher order dipolar modes ($u_1,u_2$, and $u_3$). From Eqs.~(\ref{torque}), (\ref{integral}), and (\ref{alphak}), we have calculated the integrals $I_k^{FM}, I_k^{CM}$, the coupling constants $\alpha_k^{FM},\alpha_k^{CM}$, and the torques $T^{FM}, T^{CM}$ for the FM and CM cavities. Our results are shown in Table~\ref
{tab:res2}. Note that the dominant contribution to the torque comes from the first dipolar mode, $k=1$; the higher modes give contributions of only a few per cent, at most.

\begin{table}
\caption{\label{tab:res2} Numerical Results for FM and CM cavities; $\alpha_k$ is measured in units of $(\theta/10^{-8})$ and $T_k$ is in units of $(Pb/c)(\theta/10^{-8})$}
\begin{ruledtabular}
\begin{tabular}{ccccccc}
$k$&$I^{FM}_k$&$\alpha^{FM}_k$&$T^{FM}_k$&$I^{CM}_k$&$\alpha^{CM}_k$&$T^{CM}_k$\\
\hline
1    &2.6464&0.04525 &0.33867& 2.6464& 0.00018&0.00137\\
2    &0.1136&0.00009 &0.00003& 0.1136& 0.00016&0.00005\\
3    &-0.015&-0.00000&0.00000&-0.015 &-0.0002 &0.00001\\
Total&      &        &0.33870&       &        &0.00143\\

\end{tabular}
\end{ruledtabular}
\end{table}

For mesa-beam resonators, as in the case of Gaussian-beam resonators, the nearly flat configuration (FM) is far more unstable than its nearly concentric counterpart (CM)
\begin{equation}
\frac{T^{FM}}{T^{CM}} = 237.
\end{equation}
In this case, the discrepancy is even bigger than in the Gaussian case since the eigenvalues for the FM configuration are closer to each other on the unit circle (Fig.~\ref{fig:vals}) than for the FG configuration and the phase separation of the eigenvalues for the CM configuration is even closer to $\pi$ than the phase separation of the eigenvalues for the CG configuration (Fig.~\ref{fig:vals}).

In Table~\ref{tab:tbl1}, we compare all four configurations FG, CG, FM, and CM, normalized by $T^{CG}$.
For nearly flat resonators, going from a Gaussian-beam to a mesa-beam configuration increases the strength of the instability by about a factor 5. There are two effects contributing to this increase as we can see from the following relation (in which we focus on the dominant, $k=1$ contribution):
\begin{eqnarray}
\label{torqueratio}
\frac{T^{M}}{T^{G}}=\frac{\alpha^M_1\,I^M}{\alpha^M_1\,I^M} & = & \left(\frac{\lambda^M_0+\lambda^M_1}{\lambda^G_0+\lambda^G_1}\right)\left(\frac{\lambda^G_0-\lambda^G_1}{\lambda^M_0-\lambda^M_1}\right)\left(\frac{I^M}{I^G}\right)^2 = \nonumber\\
& = & \frac{{\rm cotan}\left(\phi^{M}_{01}/2\right)}{{\rm cotan}\left(\phi^{G}_{01}/2\right)}\left(\frac{I^M}{I^G}\right)^2.
\end{eqnarray}
In the case of the nearly flat configurations both phase differences are small and since $\phi^{FM}_{01} < \phi^{FG}_{01}$ (see Fig.~\ref{fig:vals}),
\begin{equation}
\frac{{\rm cotan}\left(\phi^{FM}_{01}/2\right)}{{\rm cotan}\left(\phi^{FG}_{01}/2\right)} > 1.\nonumber
\end{equation}
This effect is amplified by the second ratio because of the higher overlap between the two eigenstates in the case of mesa beams than for Gaussian beams. This is manifested in the higher value of $I^{FM} = 2.65$ compared to $I^{FG} = 1.87$ (compare the overlaps between each pair of modes in  Fig.~\ref{fig:modes}a and  Fig.~\ref{fig:modes}b).

\begin{table}
\caption{\label{tab:tbl1}Comparison between different configurations of a fiducial optical cavity.  The torques  due to light pressure (when tilt angle $\theta$ and circulating power $P$ are the same) are normalized such that $T^{CG}=1$.}
\begin{ruledtabular}
\begin{tabular}{lll}
       &  Nearly Flat Cavity  & Nearly Concentric Cavity   \\
\hline
    G-Beam  & $T^{FG}=40.7$ & $T^{CG}=1.0$ \\
    M-Beam  & $T^{FM}=215$  & $T^{CM}=0.91$ \\
\end{tabular}
\end{ruledtabular}
\end{table}

For nearly concentric resonators, going from Gaussian-beam resonators to mesa-beam resonators weakens the net instability:  $T^{CM}/T^{CG} = 0.91$. In this case, the difference in the overlaps of the eigenstates is unchanged, but the phase differences are close to, but less than $\pi$. Since  $\phi^{CM}_{01} > \phi^{CG}_{01}$ (again look at the separation of each set of eigenvalues on the unit circle for the CG and CM configurations in Fig.~\ref{fig:vals}),
\begin{equation}
\frac{{\rm cotan}\left(\phi^{CM}_{01}/2\right)}{{\rm cotan}\left(\phi^{CG}_{01}/2\right)} < 1.\nonumber
\end{equation}
The two effects counteract each other and for this choice of parameters the net result is in favor of the CM-Beam resonator. The comparison between the torques for nearly flat and nearly concentric cavities is straightforward using Eq.~(\ref{torqueratio}) and the duality relation (see Eq.~(\ref{valsduality}) and Ref.~\cite{duality}).

In our formulation of the perturbation theory, we account for effects scaled
to first order in the tilt angle $\theta$. We assume small mode mixing
$\alpha_k \ll 1$ in order for the perturbation method to work. From our
numerical results (Table~\ref{tab:res2}), we see that  $\alpha_k \ll 1$
requires the angular orientation of the cavity mirrors be controlled to
$\theta < 10^{-8}$.\footnote{Currently, the control system of the initial LIGO interferometers operates with accuracy $\theta \simeq  10^{-7}$;  an accuracy $\theta \simeq  10^{-8}$ is planned for advanced LIGO interferometers \cite{DSigg}.}

The contributions $T_k$ of the higher order dipolar modes $k=2,3,\ldots$ to the torque can be understood by studying the analog of Eq.~(\ref{torqueratio}). From the relative locations of the eigenvalues along the unit circle and the overlapping of the eigenmodes, it is easy to show that $T_k$'s are monotonically decreasing, $T_1>T_2>T_3\ldots$. Thus, we accept the contribution from the highest dipolar mode $u_3$ in our calculation, including the numerical error, as the maximum error of the method due to neglecting the higher order dipolar modes. In this way, we conclude that the error in our total torque in the case of the CM cavity is less than 1 per cent. In the case of the FM cavity the error of the method is practically of order of the numerical error, so it is less than 0.1 per cent.

For another comparison, we perform the same calculations for the baseline design of advanced LIGO (Table~1 in \cite{baseline}). The baseline parameters were chosen such that the beam radius at the mirrors\footnote{Note that our definition for the beam radius at the mirrors differs from \cite{baseline} by factor of $\sqrt 2$.} in the case of spherical mirrors is $4.24$~cm, corresponding to diffraction losses of 10 ppm. The MH-mirror configurations are designed to have about the same diffraction losses. The resulting baseline parameters are:

$r_{max}=(15.7-0.8)$~cm $= 14.9$~cm --- the radius of the coated mirrors' surfaces.

$g^{FG}=0.9265$ --- the $g$-factor for FG resonator (corresponding mirror radius of curvature $R=54.44$~km).

$g^{CG}=-0.9265$ --- the $g$-factor for CG resonator (corresponding mirror radius of curvature $R=2.076$~km).

$r_0=b/(1-g^2)^{1/4}=4.24$~cm --- the radius of the Gaussian beam at the mirrors.

$D=3.3b=8.58$~cm --- the radius parameter of the mesa beam at the mirrors.

Table~\ref{tab:tbl2} contains the final results for these baseline parameters (including the sum of the contributions to the torques from the first three dipolar modes). Again, the least unstable configuration, and thus the easiest to control against tilt, is the nearly concentric mesa-beam (CM) resonator.

\begin{table}
\caption{\label{tab:tbl2}Comparison between different configurations of a cavity with parameters of the current baseline design for advanced LIGO.  The torques due to light pressure (when tilt angle $\theta$ and circulating power $P$ are the same) are normalized such that $T^{CG}=1$.}
\begin{ruledtabular}
\begin{tabular}{lll}
       &  Nearly Flat Cavity  & Nearly Concentric Cavity   \\
\hline
    G-Beam  & $T^{FG}=26.2$ & $T^{CG}=1.0$ \\
    M-Beam  & $T^{FM}=96$ & $T^{CM}=0.91$ \\
\end{tabular}
\end{ruledtabular}
\end{table}

\section{Conclusions}

As Table~\ref{tab:tbl2} shows, by switching from the conventional
Gaussian-beam cavities for advanced LIGO to concentric mesa-beam cavities, the
instability to symmetric tilt will be reduced (dramatically compared to a FG
cavity and moderately compared to a CG cavity). Furthermore the sensitivity of
the interferometer will improve significantly due to the reduced thermoelastic
noise (See Table~I in \cite{osv} and also \cite{tenoise1, tenoise2}).

\begin{acknowledgments}

We are very grateful to Kip Thorne who could be regarded as the `father' of this paper;
he posed the problem; it was his idea to use CM beams to reduce the tilt
instability; and he took part in very fruitful discussions and gave us useful 
advice. We thank Thorne and Yanbei Chen for helpful advice about the wording of this paper. For useful
scientific discussions, we thank Thorne, Chen, Juri Agresti, Mihai
Bondarescu, Erika d'Ambrosio, Poghos Kazarian, and Richard O'Shaughnessy. The
research reported in this paper was supported in part by National Science
Foundation grants PHY-0098715 and PHY- 0099568, by Russian Foundation
for Fundamental Research grants No.~03-02-16975-a and by contracts
No.~40.02.1.1.1.1137 and No.~40.700.12.0086 of the Industry and Science Ministry of
Russia.
\end{acknowledgments}

\appendix*
\section{Numerical Solutions of Cavity Eigenequations}\label{A}
In order to generate the set of basis solutions needed to construct perturbation theory for a cavity with arbitrary mirror shapes, we must numerically solve an integral eigenequation. We have done so using the following method, based on earlier work  by O'Shaughnessy (Sec.~V\,B of \cite{osv}).

Since the mirrors are axisymmetric [$h(\vec r) = h(r)$], we can decouple the angular and radial dependences in the eigenequations. In the numerical implementation of the eigensolver we used the following definition:

\begin{equation}
\label{angular}
u_{nm}(\vec{r})=u_{nm}(r)\,e^{-im\varphi}, \quad m=0,1,2,\ldots.
\end{equation}
Note that, for $m=0,1$, this definition of the fundamental radial mode $u_0(r)$ and the dipolar radial mode $u_1(r)$ differ from the definitions in Eqs.~(\ref{u0}) and (\ref{uk}). However, after solving the eigenequations, all modes are renormalized by numerically computing the integrals in Eqs.~(\ref{norm0}) and (\ref{normk}), so at the end we have radial modes defined as in Eqs.~(\ref{u0}) and (\ref{uk}). The resulting $u_k$ are the radial modes we need for computing $I_k$ in Eq.~(\ref{integral}). By plugging Eq.~(\ref{angular}) into Eqs.~(\ref{eigenmesa0}) and (\ref{eigenmesa1}) and integrating over the azimuthal angle, we can reduce the eigenproblem to a one-dimensional integral equation

\begin{eqnarray}
\label{eigen1d}
 \lambda_{nm}\, u_{nm}(r_1) & = & \int G_m(r_1, r_2)\, u_{nm}(r_2)\, r_2\, dr_2,\\
G_m(r_1, r_2) & = & (-i)^{m+1} J_m(r_{1}r_{2}) \nonumber \\
 & \times & {\rm exp} \left[i\left(\frac{(r_1^2+r_2^2)}{2} - 2h(r)\right)\right],\nonumber
\end{eqnarray}
where $J_m$ is the Bessel function of the first kind and order $m$.

We discretize space along the mirrors' radial direction in a uniform grid
\begin{equation}
r_j=j\,r_{max}/(N-1),\quad j=0,1,\ldots,N-1.
\end{equation}

We define the matrix $G_{(m)ij}=G_m(x_i,x_j)$, the eigenvectors $u_{(n)j}=u_n(x_j)$ ($m$, $n$ label the mode and $i$, $j$ are indices to access the matrices' and vectors' components), and we approximate the integration by a simple quadrature rule. Then the integral eigenproblem reduces to a matrix eigenvalue problem:

\begin{equation}
\label{eigenmatrix}
\lambda_{nm}\,\vec{u}_n=\hat{M}_m\,\vec{u}_n,  \textup{\; with \; } \hat{M}_{(m)ij}=\frac{r_{max}^2\,j}{N-1}G_{(m)ij}.
\end{equation}
This equation can be solved for $\lambda_{nm}$ and $\vec u_n$ by any standard matrix eigensolution software package.

\end{document}